\documentclass[showpacs,amsmath,amssymb,twocolumn,superscriptaddress,notitlepage,preprintnumbers,pra]{revtex4-1}

\usepackage[dvips]{graphicx}
\usepackage{amsmath,amssymb,amsthm,mathrsfs,amsfonts,dsfont}
 
\usepackage{dcolumn}
\usepackage{bm}
\usepackage{amsmath}
\usepackage{amssymb}
\usepackage{braket}
\usepackage{physics}
\usepackage{amsthm}
\usepackage{natbib}
\usepackage{mathtools}
\usepackage{diagbox}
\usepackage[utf8]{inputenc}
\usepackage[english]{babel}
\usepackage{scalerel}[2014/03/10]
\usepackage{stackengine}
\usepackage{algorithm}
\usepackage{relsize}
\usepackage[noend]{algpseudocode}

\usepackage{hyperref}
\hypersetup{colorlinks=true, linkcolor=blue, citecolor=blue, urlcolor=black }

\renewcommand{\tr}{\textup{Tr}}
\newcommand{\mc}[1]{\mathcal{#1}}
\newcommand{\opt}{}

\newcommand{\blu}[1]{\textcolor{black}{#1}}

\algrenewcommand\algorithmicrequire{\textbf{Input:}}
\algrenewcommand\algorithmicensure{\textbf{Output:}}

\makeatletter
\newcommand{\algmargin}{\the\ALG@thistlm}
\makeatother
\newlength{\whilewidth}
\algdef{SE}[parWHILE]{parWhile}{EndparWhile}[1]
  {\parbox[t]{\dimexpr\linewidth-\algmargin}{%
     \hangindent\whilewidth\strut\algorithmicwhile\ #1\ \algorithmicdo\strut}}{\algorithmicend\ \algorithmicwhile}%
\algnewcommand{\parState}[1]{\State%
  \parbox[t]{\dimexpr\linewidth-\algmargin}{\strut #1\strut}}

\theoremstyle{definition}

\newcommand{\PKU}{Center on Frontiers of Computing Studies, School of Computer Science, Peking University, Beijing 100871, China}

\begin{document}

\title{Experimental virtual distillation of entanglement and coherence}
\author{Ting Zhang}
\email{These authors contributed equally to this work.}
\affiliation{School of Physics, State Key Laboratory of Crystal Materials, Shandong University, Jinan 250100, China}
\author{Yukun Zhang}
\email{These authors contributed equally to this work.}
\affiliation{\PKU}

\author{Lu Liu}
\email{These authors contributed equally to this work.}
\affiliation{School of Physics, State Key Laboratory of Crystal Materials, Shandong University, Jinan 250100, China}

\author{Xiao-Xu Fang}
\affiliation{School of Physics, State Key Laboratory of Crystal Materials, Shandong University, Jinan 250100, China}

\author{Qian-Xi Zhang}
\affiliation{School of Physics, State Key Laboratory of Crystal Materials, Shandong University, Jinan 250100, China}

\author{Xiao Yuan}
\email{xiaoyuan@pku.edu.cn}
\affiliation{\PKU}

\author{He Lu}
\email{luhe@sdu.edu.cn}
\affiliation{School of Physics, State Key Laboratory of Crystal Materials, Shandong University, Jinan 250100, China}
\affiliation{Shenzhen Research Institute of Shandong University, Shenzhen 518057, China}

\begin{abstract}
Noise is in general inevitable and detrimental to practical and useful quantum communication and computation. 
Under the resource theory framework, resource distillation serves as a generic tool to overcome the effect of noise. Yet,  conventional resource distillation protocols generally require operations on multi-copies of resource states, and strong limitations exist that restrict their practical utilities. Recently, by relaxing the setting of resource distillation to only approximating the measurement statistics instead of the quantum state, a resource-frugal protocol, \textit{virtual resource distillation}, is proposed, which allows more effective distillation of noisy resources.
Here, we report its experimental implementation on a photonic quantum system for the distillation of quantum coherence (up to dimension 4) and bipartite entanglement. 
We show the virtual distillation of the maximal superposed state of dimension four from the state of dimension two, an impossible task in conventional coherence distillation. 
Furthermore, we demonstrate the virtual distillation of entanglement with operations acting only on a single copy of the noisy EPR pair and showcase the quantum teleportation task using the virtually distilled EPR pair with a significantly improved fidelity of the teleported state. These results illustrate the feasibility of the virtual resource distillation method and pave the way for accurate manipulation of quantum resources with noisy quantum hardware.
\end{abstract}

\maketitle

\vspace{0.2cm}
\noindent\textbf{\emph{Introduction.---}}A critical challenge of practical quantum information processing is to overcome the effect of noise originating from hardware imperfections. 
A systematic treatment of this problem is to exploit the language of resource theory and consider resource distillation that converts multiple copies of imperfect quantum states to the ideal one~\cite{bennett1996concentrating,vedral1997quantifying,horodecki2009quantum,horodecki2013quantumness,chitambar2019quantum}. For example, one can apply local operations and classical communication to extract a more accurate Einstein-Podolsky-Rosen (EPR) pair from multiple copies of noisy EPR pairs~\cite{Deutsch96}, generally referred to as entanglement distillation~\cite{bennett1996mixed,bennett1996purification,horodecki2009quantum}, an important procedure in long-distance quantum communication~\cite{bennett2014quantum,ekert1991quantum,kimble2008quantum,gisin2007quantum}. 
However, several theoretical works have shown strong limitations of resource distillation, which either require
many copies of the resource state or cause large distillation errors~\cite{PhysRevLett.125.060405,Regula2021fundamental,Fang2020no-go,regula_2022}. It thus remains an experimental challenge to faithfully implement accurate resource distillation for current and near-term quantum technology. 


In response to this fundamental limitation, the virtual resource distillation (VRD) protocol~\cite{yuan2023virtual,PhysRevA.109.022403} has been proposed as a promising solution. 
The key idea of VRD is to relax the conventional distillation condition to only reproduce the measurement effects (properties) of the target ideal resource state without actually synthesizing it. 
Remarkably, with proper distillation operations and classical post-processing of the measurement outcomes, the VRD scheme is able to surpass the no-go theorem of conventional resource distillations. 
For example, we could virtually obtain the ideal EPR pair using operations acting on a single copy of noisy EPR states, which is impossible conventionally.  
In Refs.~\cite{yuan2023virtual,PhysRevA.109.022403}, universal theoretical bounds have been derived for the VRD rate for general resource theories, however, these results rely on the ideal~(without implementation error) and theoretical~(practically hard to implement) assumptions of the distillation operation. Whether VRD remains practical and advantageous for existing quantum technology is still an open question. 


Here, we answer this question by presenting an experimental demonstration of virtual distillation of coherence and entanglement with an optical system. 
For quantum coherence~\cite{aberg2006quantifying,Baumgratz14,RevModPhys.89.041003}, we demonstrate the virtual preparation of a four-dimensional maximally coherent single-photon state by applying virtual distillation operations collectively on a single copy of the two-dimensional maximally coherent state, a forbidden task in conventional {coherence} resource theories~\cite{Fang2020no-go,Regula2021fundamental}. For entanglement~\cite{bennett1996concentrating,vedral1997quantifying,horodecki2009quantum}, we demonstrate {that the maximally entangled state~(MES) can be distilled from a family of two-photon Werner states~(ranging from the maximally mixed state to the MES). The distillation is confirmed by the fidelity between the virtually distilled state and the MES as well as the negativity of the virtually distilled state. Furthermore, we show that the virtually distilled state can be used for realizing quantum teleportation. We show the feasibility of VRD in the presence of experimental noise, which benefits quantum information processing using current and near-term quantum hardware. }

\vspace{0.2cm}
\noindent\textbf{\emph{Framework.---}}\label{sec:framework}
We start by briefly reviewing the framework of resource theories and virtual resource distillation~\cite{yuan2023virtual,PhysRevA.109.022403}. 
A resource theory of states consists of two basic parts: 
the set $\mc T$ of free states and the set $\mc O$ of free operations satisfying $\Gamma(\sigma)\in \mc T, \,\forall \sigma\in \mc T$ for any $\Gamma \in \mc O$. 
Conventionally, given an imperfect resource state $\rho$, the aim of resource distillation is to generate as many $m$ copies of the optimal unit pure resource state, denoted as $\psi_{\opt}$, using free operations $\Gamma\in\mc O$, such that 
\begin{equation}\label{Eq:conditionDistill}
	\frac{1}{2}\|\Gamma(\rho)-\psi^{\otimes m}_{\opt}\|_1\le \varepsilon
\end{equation}
for given accuracy $\varepsilon\ge 0$. Here $\|A\|_1 = \tr\big[\sqrt{A^\dag A}\big]$ is the trace norm. 

The idea of VRD is to first consider an equivalent form of Eq.~\eqref{Eq:conditionDistill}, i.e., $\left|\operatorname{Tr} [O \Gamma(\rho)]-\operatorname{Tr} [O \psi^{\otimes m}]\right| \leq \varepsilon$, for any bounded hermitian observable $O$. Then we relax the condition by considering linear combinations of free operations, i.e., $\tilde\Gamma = \gamma_+\Gamma_+-\gamma_-\Gamma_-$ with $\gamma_+ - \gamma_- = 1$ and $\gamma_+,\gamma_-\ge 0$,
and we have the virtual resource distillation condition
\begin{equation}\label{Eq:VconditionDistill}
	\left|\Tr \left[O \tilde\Gamma(\rho)\right]-\operatorname{Tr} \left[O \psi^{\otimes m}\right]\right| \leq \varepsilon, \forall O,
\end{equation}
which is equivalent to 
\begin{equation}\label{Eq:VconditionDistill2}
    \frac{1}{2}\big\|\tilde \Gamma(\rho)-\psi^{\otimes m}_{\opt}\|_1\le \varepsilon.
\end{equation}
Compared to conventional resource distillation where the state $\Gamma(\rho)$ is obtained, a fundamental difference here is that 
we only virtually obtain $\tilde \Gamma(\rho)$ in the sense that we can only obtain observable expectation values of $\tilde \Gamma(\rho)$. Nevertheless, we can still treat $\tilde \Gamma(\rho)$ as a virtual quantum state, and further apply quantum operations (such as state tomography~\cite{cramer2010efficient}, quantum teleportation~\cite{werner89hidden} and quantum computing) on $\tilde \Gamma(\rho)$ as long as the quantum state is finally measured.

The virtual operation $\tilde \Gamma$ could be realized in the probabilistic manner as
\begin{equation}\label{eq:vrd_prob_formula}
\begin{aligned}
    \Tr[O\tilde{\Gamma}(\rho)]=C\left(p_{+} \Tr\left[O\Gamma_{+}(\rho)\right]- p_{-} \Tr\left[O\Gamma_{-}(\rho)\right]\right)
\end{aligned}
\end{equation}
where $p_{\pm}:=\gamma_\pm/C$  and $C=\gamma_++\gamma_-$ is the cost of VRD. Specifically, we randomly apply $\Gamma_{\pm}$ to $\rho$ with probability $p_{\pm}$, measure the state with $O$, assign the minus sign when applying $\Gamma_-$, and finally multiply $C$ to the measurement outcome. Then the average of the measurement outcomes gives an unbiased estimation of $\Tr[O\tilde{\Gamma}(\rho)]$. 
Since $C\ge 1$ increases the range of the measurement outcomes, it introduces a sampling overhead, which is a common price of virtual distillation. Nevertheless, as shown in Refs.~\cite{yuan2023virtual,PhysRevA.109.022403}, even when conventional distillation is not possible, i.e., Eq.~\eqref{Eq:conditionDistill} does not satisfy for any $m\ge 1$, we could in general still run virtual resource distillation such that Eq.~\eqref{Eq:VconditionDistill} holds for nonzero $m$. We refer to Refs.~\cite{yuan2023virtual,PhysRevA.109.022403} for a detailed discussion of the virtual distillation rate and its comparison to conventional resource distillation. 

\begin{figure*}[htp]
    \centering
    \includegraphics[width=1\textwidth]{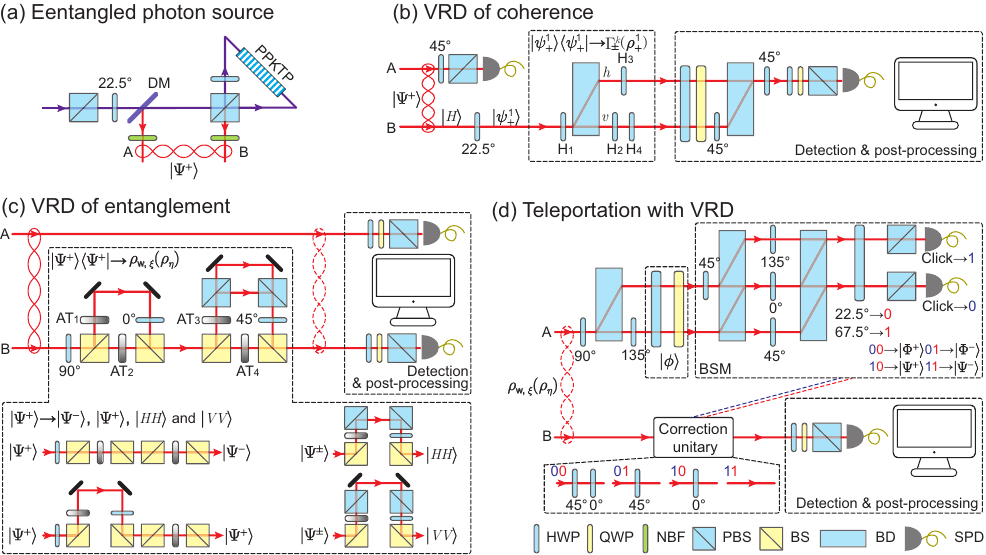}
    \caption{\blu{Schematic illustration of the experimental setup. (a)~The setup to generate polarization-entangled photon pair. (b)~The setup to demonstrate VRD of coherence on single-quarter states. (c)~Setup to generate Werner states and demonstrate VRD of entanglement on the prepared Werner states. (d)~The setup to implement quantum teleportation with assistance of VRD. The abbreviation BSM indicates Bell-state measurement. Symbols used in (a), (b), (c) and (d): half-wave plate~(HWP); quarter-wave plate~(QWP); narrow-band filter~(NBF); polarizing beam splitter~(PBS); beam splitter~(BS); beam displacer~(BD);  dichroic mirror~(DM);single-photon detector~(SPD).} Step-by-step descriptions of experimental realizations are provided in~\cite{NoteX}.}
    \label{fig:setup}
\end{figure*}

\noindent\textbf{\emph{Experiment.---}}
\label{sec:experiment}
\blu{We first generate the polarization-entangled photons with the setup shown in~Fig.~\ref{fig:setup}~(a). A periodically-poled potassium titanyl phosphate~(PPKTP) crystal is set in a Sagnac interferometer, which is bidirectionally pumped by a 405~nm ultraviolet laser diode~\cite{Kim2006PRA}. The generated two photons A and B are entangled in polarization degree of freedom~(DOF) with the ideal form of $\ket{\Psi^{+}}=\frac{1}{\sqrt{2}}(\ket{HV}+\ket{VH})$, where $H$ and $V$ represents horizontal and vertical polarization respectively. The VRD of coherence is implemented with setup shown in~Fig.~\ref{fig:setup}~(b), where photon B is detected on state $\ket{V}$ triggering photon A on state $\ket{H}$. The path DOF is manipulated by a beam displacer~(BD) with the assistance of polarization DOF, i.e., the BD transmits vertical polarization to mode $\ket{v}$ and deviates horizontal polarization to mode $\ket{h}$. We encode a single-ququart~(four-dimensional quantum system) state by $\ket{0}=\ket{H}\ket{v}$, $\ket{1}=\ket{V}\ket{v}$, $\ket{2}=\ket{H}\ket{h}$ and $\ket{3}=\ket{V}\ket{h}$. Thus, the state of photon A is $\ket{H}\ket{v}=\ket{0}$, which can be converted to $\ket{\psi_+^1}=(\ket{H}\ket{v}+\ket{V}\ket{v})/\sqrt{2}$ by applying a half-wave plate~(HWP) set at 22.5$^\circ$. Our primary goal is to virtually obtain the maximally 4-dimensional coherent state~(MCS) $\ket{\psi}_\text{MCS}=\frac{1}{2}\sum_{i=1}^4\ket{i}$ by applying virtual operations each on a single copy of the 2-dimensional coherent state $\ket{\psi_+^1}$, which is impossible in conventional resource theory~\cite{Fang2020no-go,Regula2021fundamental}. In Refs.~\cite{yuan2023virtual,PhysRevA.109.022403}, universal constructions of the VRD operations are given, however, the results assume general resource non-generation operations, which are mathematically easier yet could also be practically unphysical. Here, consider the case of $\varepsilon=0$~\footnote{However, in practice, we would always have a nonzero $\varepsilon$ due to experimental imperfections.}, we give an explicit and practical VRD protocol for input state of $\ket{\psi_+^1}$ 
\begin{equation}\label{Eq:VRDcoherence}
    \rho_\text{MCS}=C\left[p_+\Gamma_+(\rho_+^1)-p_-\Gamma_-(\rho_+^1)\right],
\end{equation}
where $\rho_\text{MCS}=\ket{\psi}_\text{MCS}\bra{\psi}_\text{MCS}$, $\rho_+^1=\ket{\psi_+^1}\bra{\psi_+^1}$, $\Gamma_+ = \sum_{k=1}^6 \Gamma_+^k$, $\Gamma_- = \sum_{k=1}^6 \Gamma_-^k$, $p_+=2/3$, $p_-=1/3$, and $C=3$. The incoherent operations $\Gamma_\pm^k$ are given in~Table~\ref{tab:GammaCoherence}, which is optimal with a minimal cost~(We refer to Supplementary Materials~\cite{NoteX} for details).} 
\begin{table}[h!b]
    \centering
    \begin{tabular}{c|cccccc}
    \hline
       $k$ & $1$ & $2$ & $3$ & $4$ & $5$ & $6$ \\
       \hline
      $\Gamma_+^k$&  $I$ & $X_{12}$ & $X_{13}$ & $X_{02}$ & $X_{03}$ & $X_{02}\oplus X_{13}$ \\
      $\Gamma_-^k$&  $Z_{01}$ & $Z_{02}X_{12}$ & $Z_{03}X_{13}$ & $Z_{12}X_{02}$ & $Z_{13}X_{03}$ & $Z_{23}(X_{02}\oplus X_{13})$ \\
    \hline
    \end{tabular}
    \caption{The VRD operations for quantum coherence. Here $X_{jk} = \ket{j}\bra{k}+\ket{k}\bra{j}$ and $Z_{jk}=\ket{j}\bra{j}-\ket{k}\bra{k}$, which are all incoherent operations~\cite{Baumgratz14}. The operations $\Gamma_\pm^k$ correspond to the unitary in the table.}
    \label{tab:GammaCoherence}
\end{table}

\blu{Experimentally, four HWPs~(H$_1\sim$ H$_4$ in~Fig.~\ref{fig:setup}~(b)) with angle setting of $\bm{\theta}_\pm^k=[\theta_{k_1}, \theta_{k_2}, \theta_{k_3}, \theta_{k_4}]$ are used to realize the 12 individual incoherent operations $\Gamma_\pm^k$~(see Supplementary Materials~\cite{NoteX} for details of $\bm{\theta}_\pm^k$). In our experiment, the 12 incoherent operations are applied on $\rho_+^1$ with equal probability, each of which leads the outcome state $\rho_\pm^k=\ket{\psi_\pm^k}\bra{\psi_\pm^k}=\Gamma^k_\pm(\rho_+^1)$. For each $\rho_\pm^k$, we perform Pauli measurements on polarization DOF and path DOF respectively. Then, we estimate the density matrix $\hat{\rho}_{\pm}^{k, \text{LIN}}$ using linear inversion~(LIN) and calculate $\hat{\rho}_\text{MCS}^\text{LIN}$ according to~Eq.~\eqref{Eq:VRDcoherence}. Note that $\hat{\rho}_\text{MCS}^\text{LIN}$ is not a physical state as it generally has negative eigenvalues. This issue can be addressed by projecting $\hat{\rho}_\text{MCS}^\text{LIN}$ onto the set of physical states with respect to the Frobenius distance, which returns a physical state $\hat{\rho}_\text{MCS}$. Hereafter, $\hat{\rho}^\text{LIN}$ represents the estimation with LIN and $\hat{\rho}$ represents the corresponding physical state obtained from $\hat{\rho}^\text{LIN}$ using projective method.  }

\begin{figure}[h]
     \centering
     \includegraphics[width=0.48\textwidth]{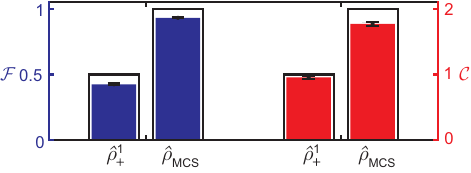}
     \caption{The experimental results of VRD of quantum coherence. Two figures of merit are adopted to indicate the improvements of distilled state $\hat{\rho}_\text{MCS}$, i.e., the fidelity $\mathcal {F}$~(blue bars) and relative entropy of coherence $\mathcal{C}$~(red bars). The black boxes represent the theoretical values. The statistic error is calculated from Poissonian counting statistics of the raw detection events. }
     \label{fig:coherence}
 \end{figure}

\blu{The distilled state $\hat{\rho}_\text{MCS}$ is firstly characterized by calculating its fidelity with respect to ideal $\rho_\text{MCS}$, i.e., $\mathcal F(\hat{\rho}_\text{MCS}, \rho_\text{MCS})=\tr(\hat{\rho}_\text{MCS}\rho_\text{MCS})$~\cite{Liang2019RPP}. The result is shown in Fig.~\ref{fig:coherence}, and we observe $\mathcal F(\hat{\rho}_\text{MCS}, \rho_\text{MCS})=0.932\pm 0.004$. We also calculate the fidelity between input state $\hat{\rho}_{+}^1$ and $\rho_\text{MCS}$, and observe $\mathcal F(\hat{\rho}_{+}^1, \rho_{\text{MCS}})=0.426\pm0.007$. The enhancement of fidelity evidently convinces the successful distillation of $\ket{\psi_{\text{MCS}}}$. The distillation is further confirmed by calculating the relative entropy of coherence~$\mathcal{C}(\rho)=\tr(\rho\log_2\rho)-\tr(\rho_d\log_2\rho_d)$ of $\hat{\rho}_{+}^1$ and $\hat{\rho}_\text{MCS}$ respectively, where $\rho_d$ is the diagonal matrix of $\rho$~\cite{BaumgratzPRL2014}. We observe that $\mathcal{C}(\hat{\rho}_\text{MCS})=1.769\pm0.029$ while $\mathcal{C}(\hat{\rho}_{+}^1)=0.958\pm0.024$ as shown in Fig.~\ref{fig:coherence}.} Both the fidelity $\mathcal{F}$ and the relative entropy of coherence $\mathcal{C}$ are significantly enhanced with the VRD protocol.

\begin{figure*}[htp]
\includegraphics[width=0.98\textwidth]{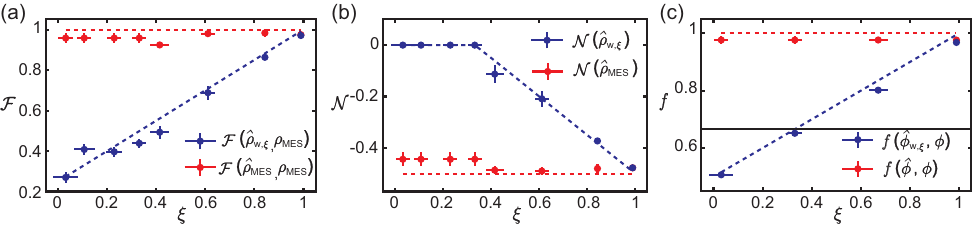}
   \caption{Experimental results of VRD of entanglement. 
   (a)~Fidelity of $\mathcal{F}(\hat{\rho}_\text{MES}, \rho_\text{MES})$ and $\mathcal{F}(\hat{\rho}_{\text{w}, \xi}, \rho_\text{MES})$. (b)~Negativity of $\mathcal{N}(\hat{\rho}_\text{MES})$ and $\mathcal{N}(\hat{\rho}_{\text{w}, \xi})$. (c)~The average fidelity $f$ of quantum teleportation with 
   $\hat{\phi}_{\text{w}, \xi}$ or $\hat{\phi}$. Dashed lines represents the theoretical predictions with the ideal states, and dots represents the experimental results. The blue color represents results before VRD and the red color represents the results after VRD. The black line in (c) represent classical threshold $f_c=2/3$ in quantum teleportation.}
    \label{fig:ent}
\end{figure*}

Next, we implement the VRD of MES $\ket{\Psi^{-}}=\frac{1}{\sqrt{2}}(\ket{01}-\ket{10})$ from a family of two-qubit Werner states in the form of~\cite{Werner89} 
\begin{equation}\label{Eq:WS}
\rho_{\text{w}, \xi}=\xi\Psi^{-} +(1-\xi) \mathds1_4 / 4, 0\leq \xi\leq1
\end{equation}
with $\Psi^{-}=\ket{\Psi^{-}}\bra{\Psi^{-}}$. Note that the two-qubit Werner states are separable for $\xi< 1/3$ and entangled for $\xi\geq 1/3$ according to the positive partial transpose (PPT) criterion~\cite{peres1996separability,horodecki2001separability}. 
The VRD protocol for Werner states is given by
\begin{equation}\label{eq:ent_proto}
   \rho_\text{MES}=\Psi^{-}=C_\xi\left[p_{+} \Gamma_{+}\left(\rho_{\text{w}, \xi}\right)- p_{-} \Gamma_{-}\left(\rho_{\text{w}, \xi}\right)\right],
\end{equation}
where $\Gamma_+$ is the identity operation,
$\Gamma_{-}(\cdot)=\rho_\eta=(\mathds1_4-\Psi^{-})/3$ is a replacement channel,  $p_+=\frac{4}{7-3\xi}$, $p_-=\frac{3-3\xi}{7-3\xi}$ and $C_\xi=\frac{7-3\xi}{1+3\xi}$. In the cases $\xi< 1/3$, the VRD is performed with input state $\rho_{\text{w}, 1/3}$ regardless of $\xi$.~(see Supplementary Materials~\cite{NoteX} for the proof of optimality of this VRD protocol.)

\blu{Experimentally, the Werner state is encoded in polarization DOF, i.e., $\ket{H}=\ket{0}$ and $\ket{V}=\ket{1}$. The Werner states in~Eq.~\eqref{Eq:WS} can be rewritten by a mixture of four pure states
\begin{equation}\label{Eq:WSexp}
\rho_{\text{w}, \xi}=\frac{1+3\xi}{4}\Psi^-+\frac{1-\xi}{4}\left(\Psi^++\rho_{HH}+\rho_{VV}\right), 
\end{equation} 
where $\rho_{HH}=\ket{HH}\bra{HH}$ and $\rho_{VV}=\ket{VV}\bra{VV}$. As shown in~Fig.~\ref{fig:setup}~(c),  $\rho_\xi^\text{w}$ can be prepared with an array of beam splitters~(BSs), polarizing beam splitters~(PBSs), HWPs and attenuators~(ATs). Specifically, the setup is able to covert input state $\ket{\Psi^+}$ to a mixture of $\Psi^-$, $\Psi^+$, $\rho_{HH}$ and $\rho_{VV}$, where the probability of each component is determined by the transmittance of ATs. Thus, the parameter $\xi$ is determined by $\bm{t}_\xi=[t_{\xi_1}=\frac{1-\xi}{2+2\xi}, t_{\xi_2}=\frac{1+3\xi}{2+2\xi}, t_{\xi_3}=\frac{1-\xi}{2}, t_{\xi_4}=\frac{1+\xi}{2}]$ with $t_{\xi_i}$ being the transmittance of the $i$th AT~(See Supplementary Materials~\cite{NoteX} for more details). In our experiment, we prepare eight $\rho_{\text{w}, \xi}$ with parameter $\xi\in[0, 0.1, 0.2, 1/3, 0.4, 0.6, 0.8, 1]$ for distillation. For each $\xi$, the VRD requires the preparation of $\rho_\eta$, which can be decomposed into $\rho_\eta=\frac{1}{3}\left(\Psi^++\rho_{HH}+\rho_{VV}\right)$ and can be prepared with the same setup by setting $\bm{t}_\eta=[t_{\eta_1}=1, t_{\eta_2}=0, t_{\eta_3}=\frac{2}{3}, t_{\eta_4}=\frac{1}{3}]$. We set the parameters of four ATs being $\bm{t}_\xi$ and $\bm{t}_\eta$ with equal probability, and estimate the corresponding density matrix $\hat{\rho}^\text{LIN}_{\text{w}, \xi}$ and $\hat{\rho}_\eta^\text{LIN}$.  Then, $\hat{\rho}_\text{MES}^\text{LIN}$ is calculated according to Eq.~\eqref{eq:ent_proto}, and $\hat{\rho}_\text{MES}$ is consequently obtained.  }

\blu{We calculate the fidelity $\mathcal{F}(\hat{\rho}_\text{MES}, \rho_\text{MES})$ and $\mathcal{F}(\hat{\rho}_{\text{w}, \xi}, \rho_\text{MES})$. As shown in~Fig.~\ref{fig:ent}~(a), we observe that $\mathcal{F}(\hat{\rho}_\text{MES}, \rho_\text{MES})$ is close to 1 regardless of the value of $\xi$ while $\mathcal{F}(\hat{\rho}_{\text{w}, \xi}, \rho_\text{MES})$ is linearly dependent on the value of $\xi$.} To further investigate the quantity of entanglement, we calculate the entanglement negativity of a quantum state~\cite{vidal2002computable}
\begin{equation}
    \mathcal{N}(\rho)= \frac{\left\|\rho^{T_\text{A}}\right\|_1-1}{2},
\end{equation}
where $T_\text{A}$ represents partial transpose of the density matrix $\rho$ with respect to its subsystem A and $\|\cdot\|_1$ is the trace norm. Note that $\mathcal N=0$ for separable states and $\mathcal N=-0.5$ for MES. The negativity of $\hat{\rho}_{\text{w}, \xi}$ and $\hat{\rho}_\text{MES}$ are shown in~Fig.~\ref{fig:ent}~(b), in which we observe that the amount of entanglement is significantly enhanced after distillation. In particular, $\hat{\rho}_{\text{w}, \xi}$ with $\xi=0, 0.1, 0.2$ and $1/3$ are separable states as their negativities are 0, which indicates no entanglement exists in these states. However, the corresponding virtually distilled states admit a large amount of entanglement as their negativities are close to $-0.5$~\footnote{We note that this seems controversial since we obtained entanglement from separable states. Indeed, this becomes possible as we allow classical post-processing in resource distillation and the entanglement is only virtually/effectively created. There also exists a sampling overhead, which would exponentially increase with $n$ if we consider the virtual distillation of $n$ maximally entangled states from separable states.}.

\blu{Finally, we demonstrate quantum teleportation with assistance of VRD. $\phi=\ket{\phi}\bra{\phi}$ is the quantum state to be teleported, and the teleportation is performed with $\rho_{\text{w}, \xi}$ and its corresponding $\rho_\eta$ respectively. After teleportation, the output state is denoted by $\phi_{\text{w}, \xi}$ and $\phi_\eta$ respectively, and  the VRD for teleportation is 
\begin{equation}
\phi=\frac{4}{1+3\xi}\phi_{\text{w}, \xi}-\frac{3-3\xi}{1+3\xi}\phi_\eta.
\end{equation}
Experimentally, we implement the two-photon teleportation scheme~\cite{Boschi1998PRL,jin2010experimental,jiang2020quantum,LiJyun-Yi21} as illustrated in~Fig.~\ref{fig:setup}~(d). We prepare $\rho_{\text{w}, \xi}$ and $\rho_\eta$ with $\xi\in[0,1/3,2/3,1]$, and select four states encoded in polarization DOF for teleportation, i.e., $\ket{\phi}\in[\ket{H}, \ket{V}, \ket{+}=\frac{1}{\sqrt{2}}(\ket{H}+\ket{V}), \ket{R}=\frac{1}{\sqrt{2}}(\ket{H}+i\ket{V}]$. We calculate the average fidelity $f(\hat{\phi}, \phi)=\frac{1}{4}\sum_\phi \mathcal F(\hat{\phi}, \phi)$ and $f(\hat{\phi}_{\text{w},\xi}, \phi)=\frac{1}{4}\sum_\phi \mathcal F(\hat{\phi}_{\text{w},\xi}, \phi)$, where $\hat{\phi}$ and $\hat{\phi}_{\text{w},\xi}$ are the reconstructed output states with VRD protocol and $\rho_{\text{w}, \xi}$ respectively.  As shown in Fig.~\ref{fig:ent}(c), $f(\hat{\phi}, \phi)$ is close to the ideal value of 1~(teleportation with maximal entanglement) while $f(\hat{\phi}_{\text{w},\xi}, \phi)$ is strongly related to $\xi$, which is similar to the results in Fig.~\ref{fig:ent}(a).} In particular, $f(\hat{\phi}_{\text{w},\xi}, \phi)$ is below $f_c=2/3$ when $\xi=0$, where $f_c=2/3$ is the classical limit in quantum teleportation~\cite{Horodecki99general}. The experimental results indicate the practical usefulness of the VRD method is able to circumvent the theoretical barrier~\cite{yuan2023virtual,PhysRevA.109.022403} of conventional distillation methods where the distillation may not be feasible for $\xi\neq1$.

\vspace{0.2cm}
\noindent\textbf{\emph{Conclusion.---}}In this work, we experimentally study the performance of the VRD protocol for coherence and entanglement. Besides, we demonstrate quantum teleportation with VRD. In all three experiments, we see that despite imperfect quantum operations involved in the distillation process, our results show significant improvements after the distillation. Moreover, we demonstrate the distinct features of the VRD protocol for circumventing obstacles for conventional distillation schemes. We thus expect the VRD protocol to find more utilities in practical situations such as quantum communication~\cite{bennett2014quantum,ekert1991quantum,kimble2008quantum,gisin2007quantum} and quantum networks~\cite{cirac1997quantum,duan2001long,wehner2018quantum}.
 
Regarding future work, it is crucial to extend the VRD method to a more realistic scenario such that the noisy resource state is unknown beforehand. Furthermore, the performance of the VRD protocol is affected by the knowledge of the noise channel. As a proof of principle, we manufacture the imperfect state in this work. Yet, as the noisy state is not easily accessible in a realistic scenario, it is not obvious how the VRD protocol can be applied in a resource-efficient way at first glance. We remark that the unknown noisy state can be approximated through the estimation of the noise channel using statistical machine learning frameworks~\cite{harper2020efficient,pastori2022characterization,van2023probabilistic,kim2023evidence,rouze2023efficient}. It is worth noting that several experiments have adopted the noisy-channel learning methods as a subroutine for implementing quantum error mitigation algorithms~\cite{van2023probabilistic,kim2023evidence}. The combination of these learning techniques with the VRD protocol may be fruitful in extending the utility of the VRD method.
Finally, although virtual distillation outperforms conventional resource distillation, it is not the optimal way of processing information, see Supplementary Materials~\cite{NoteX} for details. 
Thus, it is interesting to study other protocols to more efficiently retrieve information from noisy resources.

\vspace{.1cm}

\begin{acknowledgments}
\noindent{\emph{Acknowledgments.---}}
We thank the anonymous referees for the insightful comments on the work, which helped clarify the advantages and limitations of virtual resource distillation. This work is supported by the National Key R\&D Program of China (Grant No.~2019YFA0308200), the National Natural Science Foundation of China  (Grants No.~11974213, No.~92065112, No.~12175003, and No.~12361161602), NSAF Grant No.~U2330201, Shandong Provincial Natural Science Foundation (Grants
No.~ZR2020JQ05), Taishan Scholar of Shandong Province (Grant No.~tsqn202103013), Shenzhen Fundamental Research Program (Grant No.~JCYJ20190806155211142), Shandong University Multidisciplinary Research and Innovation Team of Young Scholars (Grant No.~2020QNQT), the Higher Education Discipline Innovation Project (`111') (Grant No.~B13029) and the High-performance Computing Platform of Peking University.
\end{acknowledgments}

\bibliography{main}

\appendix
\onecolumngrid

\section{Theory}
In the context of quantum resource theory, conventional distillation protocol focuses on utilizing `free operations' to transform a noisy state into a less noisy one. Despite powerful results established over the years, fundamental limitations are found for highly mixed input states.
To this end, \textcite{yuan2023virtual,PhysRevA.109.022403} proposed the virtual resource distillation (VRD) protocol~\cite{yuan2023virtual,PhysRevA.109.022403} that generalizes the conventional distillation settings that allow the purification of states more flexibly by reproducing the measurement statistics of the purified state.

The main focus of this work is to demonstrate the practical utility of VRD protocol~\cite{yuan2023virtual,PhysRevA.109.022403} experimentally. As we are in the noisy intermediate-scale quantum (NISQ) era~\cite{preskill2018quantum}, near-term quantum devices are prone to all kinds of noise such that the performance of this quantum hardware is compromised drastically as the system size increases. The conventional resource distillation protocol aims at obtaining a less noisy (purified) state from several copies of noise input states. By contrast, the VRD method tries to reproduce the measurement outcome of the ideal quantum process through methods such as Monte Carlo simulation approaches~\cite{pashayan2015estimating,seddon2021quantifying,peng2020simulating,bravyi2022simulate}, quantum error mitigation~\cite{temme2017error,li2017efficient,endo2021hybrid} and unphysical process~\cite{buscemi2013direct,jiang2021physical,regula2021operational}.
Consequently, the expectation values of observables (measurement outcomes) are improved such that they become closer to the ideal outcomes. The variance of the measurement outcome could be enlarged, and so is the sample complexity of the VRD protocol. Consequently, the efficiency of distillation of the VRD protocol may be less favorable when compared to traditional distillation methods.
However, as the VRD approach only requires one copy of the input state and still works even when the input state is highly mixed, these features allow the VRD protocol to surpass several conventional limitations as discussed in Refs.~\cite{yuan2023virtual,PhysRevA.109.022403}.

Crucially, the central focus of the VRD protocol is concentrated on the derivation of VRD schemes that achieve as much efficiency as possible. To this end, the one-shot VRD rate is introduced~\cite{yuan2023virtual,PhysRevA.109.022403} to describe the distillation efficiency and it is found that the optimal result is characterized by a semi-definite programming (SDP) problem~\cite{mironowicz2023semi}. It is known that the SDP problem can be solved efficiently by a classical computer. As such, the optimal VRD scheme can be derived as long as the one-shot VRD rate.

In the following, we will provide details for the derivation of the distillation of coherence and entanglement.

\subsection{Framework for virtual resource distillation}
\label{sec:frame}
In this section, we introduce the framework and key details of the VRD protocol. As introduced in the main text, the two basic parts of a resource theory are the set $\tau$ of free states and the set $\mathcal{O}$ of free operations satisfying $\Gamma(\sigma) \in \tau, \forall\sigma\in \tau$ for any $\Gamma\in \mathcal{O}$.
For example, we can consider the set of operations comprised of resource non-generating (RNG) operations, i.e. free operations $\Gamma$ such that when acting on free states we will obtain another free state. 

In the traditional setting of resource theory, the aim is to reproduce $k$ copies of an optimal unit pure resource state $\psi$ from imperfect density matrix $\rho$ within given accuracy $\epsilon$ in trace distance. For the VRD setting, the optimal unit pure resource state is no longer required. As discussed in the main text, we loosen this condition by allowing reproducing information from $\psi$ with the given density matrix $\rho$. The information reproduction is achieved by retrieving the expectation value of given observable (more generally, measurement outcome) $O$ regarding $\psi$. Formally, the VRD tasks can be stated as follows.
For any hermitian operator $O$ such that $0\preccurlyeq O \preccurlyeq I$ and given accuracy $\epsilon\in (0,1)$, the VRD task is to achieve
\begin{equation}
    \left|\operatorname{Tr} O \Gamma(\rho)-\operatorname{Tr} O \psi^{\otimes k}\right| \leq \epsilon.
\end{equation}
The above definition provides extra flexibility over the conventional setting as the information we aim to reproduce is a classical measurement outcome so that a linear combination of free operations is allowed. Explicitly, we can expand $\Gamma(\cdot)$ as linear combination of RNG operations
\begin{equation}\label{eq:free_op_decompositon}
    \sum_j \gamma_j \operatorname{Tr}\left(M \Gamma_j(\rho)\right)=\operatorname{Tr}\left(M \sum_j \gamma_j \Gamma_j(\rho)\right),
\end{equation}
where coefficients $\gamma_j$ satisfies $\sum_j \gamma_j=1$. 
Eq.~\eqref{eq:free_op_decompositon} is then effectively achieved by first grouping operations with the same sign together
\begin{equation}
    \left|\operatorname{Tr}\left[O\left(\gamma_{+} \Gamma_{+}(\rho)-\gamma_{-} \Gamma_{-}(\rho)\right)\right]-\operatorname{Tr} O \psi^{\otimes m}\right| \leq \epsilon,
\end{equation}
where $\gamma_{ \pm}:=\sum_{j: \operatorname{sign}(j)= \pm 1} \gamma_j \geq 0$, $\gamma_{+}-\gamma_{-}=1$ and $\Gamma_{\pm}:=\frac{1}{\gamma_{\pm}} \sum_{j: \operatorname{sign}(j)= \pm 1} \gamma_j \Gamma_j$. Then, we write the operation in a probabilistic manner
\begin{equation}\label{eq:gamma_expand}
    \Tr[O\tilde{\Gamma}(\rho)]=\operatorname{Tr}\left[O\left(\gamma_{+} \Gamma_{+}(\rho)-\gamma_{-} \Gamma_{-}(\rho)\right)\right]=C\operatorname{Tr}\left[O\left(p_{+} \Gamma_{+}(\rho)- p_{-} \Gamma_{-}(\rho)\right)\right],
\end{equation}
where $p_{\pm}:=\gamma_\pm/C$ is the probability to implement $\Gamma_\pm(\cdot)$ to the noisy input state, and $C:=\gamma_{+}+\gamma_{-}$ serves as the normalization factor of the probabilistic distribution. By Hoeffding's inequality~\cite{hoeffding1994probability}, an extra $C^2$-fold cost is introduced in sample complexity. As we only consider one copy of the input state, this cost is reflected in the one-shot distillation rate. 
More formally, Refs.~\cite{yuan2023virtual,PhysRevA.109.022403} employs the one-shot VRD rate to characterize the cost. That is
\begin{equation}\label{eq:vrd_formula}
    V^{\varepsilon}(\rho)=\sup _m \frac{m}{C^{\varepsilon}(\rho, m)^2},
\end{equation}
and the normalization factor is
\begin{equation}
    C^{\varepsilon}(\rho, m)=\inf _{\substack{\tilde{\Gamma}=\gamma_{+}\Gamma_{+}-\gamma_{-}\Gamma_{-} \\ \gamma_{+}-\gamma_{-}=1 \\ \Gamma_{ \pm} \in \Gamma, \gamma_{ \pm} \geq 0}}\left\{\gamma_{+}+\gamma_{-}: \frac{1}{2}\left\|\tilde{\Gamma}(\rho)-\psi^{\otimes m}\right\|_1 \leq \epsilon\right\},
\end{equation}
where $\|A\|_1=\Tr\sqrt{A^\dagger A}$ is the trace distance.


\subsection{Coherence}
In this section, we apply the VRD framework to the task of coherence distillation. Our aim is to obtain the maximally coherent state $\ket{\psi}_\text{MCS}=\frac{1}{2}\sum_{i=1}^4\ket{i}$ from the partially coherent state $\ket{\psi_+^1}=\frac{1}{2}(\ket{0}+\ket{1})$ with free operations using the VRD method. 

\begin{table}[b]
    \centering
    \begin{tabular}{c|cccccc}
    \hline
       $k$ & $1$ & $2$ & $3$ & $4$ & $5$ & $6$ \\
       \hline
      $\Gamma_+^k$&  $I$ & $X_{12}$ & $X_{13}$ & $X_{02}$ & $X_{03}$ & $X_{02}\oplus X_{13}$ \\
      $\ket{\psi_+^k}$&  $\frac{1}{2}(\ket{0}+\ket{1})$ & $\frac{1}{2}(\ket{0}+\ket{2})$ & $\frac{1}{2}(\ket{0}+\ket{3})$ & $\frac{1}{2}(\ket{1}+\ket{2})$ & $\frac{1}{2}(\ket{1}+\ket{3})$ & $\frac{1}{2}(\ket{2}+\ket{3})$ \\
      $\Gamma_-^k$&  $Z_{01}$ & $Z_{02}X_{12}$ & $Z_{03}X_{13}$ & $Z_{12}X_{02}$ & $Z_{13}X_{03}$ & $Z_{23}(X_{02}\oplus X_{13})$ \\
      $\ket{\psi_-^k}$&  $\frac{1}{2}(\ket{0}-\ket{1})$ & $\frac{1}{2}(\ket{0}-\ket{2})$ & $\frac{1}{2}(\ket{0}-\ket{3})$ & $\frac{1}{2}(\ket{1}-\ket{2})$ & $\frac{1}{2}(\ket{1}+\ket{3})$ & $\frac{1}{2}(\ket{2}-\ket{3})$ \\
    \hline
    \end{tabular}
    \caption{The VRD operations and output states for quantum coherence. Here $X_{jk} = \ket{j}\bra{k}+\ket{k}\bra{j}$ and $Z_{jk}=\ket{j}\bra{j}-\ket{k}\bra{k}$, which are all incoherent operations~\cite{Baumgratz14}.}
    \label{tab:Coherence}
\end{table}

In the main text, we have prescribed the set of states that we need to prepare by applying free operations to the input state. Here, we show that the derived operations (prepared states) are optimal. The free operations are obtained by solving the following optimization problem:
\begin{equation}\label{eq:optimal}
\begin{aligned}
& \zeta_{\varepsilon}^s(\rho, k):=\min \gamma_{+}+\gamma_{-} \\
& \text {s.t. } 0 \leq Q_{+} \leq \gamma_{+} I, 0 \leq Q_{-} \leq \gamma_{-} I, \\
& \operatorname{Tr} Q_{+} \sigma \leq \frac{\gamma_{+}}{k}, \operatorname{Tr} Q_{-} \sigma \leq \frac{\gamma_{-}}{k}, \forall \sigma \in \mathcal{\tau}, \\
& \gamma_{+}-\gamma_{-}=1, \operatorname{Tr} \rho\left(Q_{+}-Q_{-}\right) \geq 1-\varepsilon,
\end{aligned}
\end{equation}
where $k$ is the parameter. Another optimization problem $\zeta_{\varepsilon}^g(\rho, k)$ is defined such that we modify the constraints in the second line to be $\operatorname{Tr} Q_{+} \gamma=\mu_{+} / k, \quad \operatorname{Tr} Q_{-} \gamma=\mu_{-} / k, \forall \gamma \in \tau$ with different choices of $k$. Crucially, solutions to the two optimization problems $\zeta_{\varepsilon}^{s/g}(\rho, k)$ allow us to upper and lower bound the cost of the VRD method as shown in [Theorem 1, Refs.~\cite{yuan2023virtual,PhysRevA.109.022403}].
We remark that the optimization problem defined by $\zeta_{\varepsilon}^s(\rho, k)$ is efficiently solvable by a classical computer using positive semi-definite programming (SDP) approaches. Moreover, in the cases where these two bounds coincide, the optimal solution can be exactly determined. This condition holds in the scenarios of coherence and entanglement distillation. And, the solution is given by 
\begin{equation}\label{eq:exact_cost}
    C^{\varepsilon}(\rho, m)=\zeta_{\varepsilon}^{s / g}\left(\rho, F_{\tau}\left(\psi^{\otimes m}\right)^{-1}\right),
\end{equation}
where $F_{\tau}(\rho)=\max _{\sigma \in \tau}\left(\operatorname{Tr} \sqrt{\rho^{1 / 2} \sigma \rho^{1 / 2}}\right)^2$ is the measure of resource fidelity.
Therefore, the optimality of the one-shot VRD rate that relates to the cost of the distillation is justified by the tightness of the bounds. By solving the SDP problems,
we obtain the operation as 
\begin{equation}\label{eq:coherence_formula}
\begin{aligned}
    \tilde{\Gamma}(\rho_+^1)&=\gamma_+\Gamma_+(\rho_+^1)-\gamma_-\Gamma_+(\rho_-^1)\\
    &=C^0\left(\sum_{k=1}^6 p^k_+ \Gamma_+^k(\rho_+^1) - \sum_{k=1}^6 p^k_- \Gamma_-^k(\rho_+^1) \right),
\end{aligned}
\end{equation}
where $\rho^1_+=\ket{\psi_+^1}\bra{\psi_+^1}$, $\gamma_+=2$, $\gamma_-=1$, $C^0=\gamma_++\gamma_-$, and $p^k_+=\frac{2}{3},~p^k_-=\frac{1}{3} \forall k$, respectively. Let us explain. In the first line of Eq.~\eqref{eq:coherence_formula}, we expand the channel $\tilde{\Gamma}$ into two parts as shown in Eq.~\eqref{eq:gamma_expand}. Then, in the second line, we normalize the coefficients of the two sets of operations such that each operation $\Gamma_{+/-}^k$ consists $\Gamma_{+/-}$ is implemented probabilistically with $p^k_{+/-}$. It is clear to see that the cost $C^0$ is the normalization factor that satisfies $C^0>1$ so that the variance for the calculation is enlarged. To show the formula given in Eq.~\eqref{eq:coherence_formula} indeed provides a feasible VRD protocol, we specify explicit forms of each operation $\Gamma_{+/-}^k$ along with the output state $\Gamma_{+/-}^k(\rho^1_+)$ in Table \ref{tab:Coherence}. Here, because the ideal output states are all pure, we provide them in the corresponding form $\ket{\psi_{+/-}^k}$. One can readily check from the table that all operations $\Gamma_{+/-}^k$ are resource non-generative such that the coherence of the output state is no larger than the input state. Moreover, the validity of the distillation is justified by the fact that the target density matrix of $\ket{\psi}_\text{MCS}=\frac{1}{2}\sum_{i=1}^4\ket{i}$ is produced by the operations given by Eq.~\eqref{eq:coherence_formula}.

\subsection{Entanglement distillation}
In this section, we provide details regarding the distillation of entanglement and the applicability of the VRD protocol in quantum teleportation. We will show that although the VRD method reproduces the statistics of the ideal state, not the state itself, applications are fitted naturally into the VRD framework if the expectation values of observables are interested. The target is to distill the maximally entangled state $\ket{\Psi^{-}}$ from the noisy input state, which is the classes of Werner states in the following form
\begin{equation}\label{eq:WS}
\rho_{\text{w}, \xi}=\xi\ket{\Psi^{-}}\bra{\Psi^{-}}+(1-\xi) \mathds1_4 / 4, 0\leq \xi\leq1.
\end{equation}

The positive partial transpose (PPT) criterion~\cite{peres1996separability,horodecki2001separability} indicates that the two-qubit Werner states are separable for $\xi< 1/3$ and entangled for $\xi\geq 1/3$. We take two different strategies under these two conditions as follows. When $\xi\geq 1/3$, the VRD protocol is defined as 
\begin{equation}\label{eq:ent_proto}
    \Psi^-=\tilde{\Gamma}\left(\rho_{\text{w}, \xi}\right)=\frac{4}{1+3 \xi} \Gamma_{+}\left(\rho_{\text{w}, \xi}\right)-\frac{3-3 \xi}{1+3 \xi} \Gamma_{-}\left(\rho_{\text{w}, \xi}\right),
\end{equation}
where $\Gamma_+(\rho):=\rho$ is the identity channel and $\Gamma_{-}(\rho)=\rho_\eta:=(\mathds1_4-\Psi^{-})/3$. The latter channel replaces any input state with $\rho_\eta$. 
While in cases $\xi< 1/3$, the VRD protocol~\cite{yuan2023virtual} proposed to replace the density matrix $\rho_{\text{w}, \xi}$ with $\rho_{\text{w}, 1/3}$, it is worth noting that the distillation channel gives by Eq.~\eqref{eq:ent_proto} still works. The main reason for applying the replacement channel is to maintain the VRD protocol's optimality as provided by solving the SDP problem defined in Eq.~\eqref{eq:optimal}. Ideally, we assume that the free operations involved in $\tilde{\Gamma}$ can be implemented error-free. In such scenarios, the VRD protocol can perfectly reproduce the target state, i.e.~$\varepsilon=0$. Thus, taking the error $\varepsilon=0$, we get the overhead
\begin{equation}\label{eq:ent_cost}
    C^0\left(\rho_{\text{w}, \xi}\right)=\min \{(7-3\xi)/(1+3\xi), 3 \}.
\end{equation}
In both cases, we obtain the channel $\tilde{\Gamma}$ by solving the SDP problem defined in Eq.~\eqref{eq:optimal}. The feasibility of the two different distillation schemes can be checked by the fact that the output state of Eq.~\eqref{eq:ent_proto} is indeed the targeted maximally entangled state. Moreover, the optimality of the solution is promised as the upper and lower bound on the cost coincide in the case of entanglement distillation such that the estimated cost provided in Eq.~\eqref{eq:exact_cost} becomes exact.

We then utilize the distilled state for the task of quantum teleportation. As the process of quantum teleportation is affected by noises, traditionally, one employs the entanglement distillation scheme to distill one copy of a purified state from several copies of noisy states. Yet, the VRD protocol provides extra flexibility to the process of distillation and allows us to obtain a better result with only a single copy of the noisy state. To see this, let us consider the scenario that Alice
wants to send a state $\phi_\text{C}=\ket{\phi}\bra{\phi}_\text{C}$ to Bob with the assistance of the maximally entangled state shared between  Alice and Bob $\ket{\Psi^{-}}_\text{AB}=\frac{1}{\sqrt{2}}\left(\ket{01}_\text{AB}-\ket{10}_\text{AB}\right)$. To this end, Alice performs Bell state measurement~(BSM) on her qubits A and C, then sends the results of BSM to Bob. Depending on the results of the BSM, Bob implements correction unitary operation on his qubit B to recover $\phi_\text{B}$. The correction unitary operations depending on the result of BSM are: $00\to ZX$, $01\to X$, $10\to Z$, and $11\to \mathds 1$.

The VRD for teleportation is 
\begin{equation}
\phi_\text{B}=\frac{4}{1+3\xi}\phi_{\text{w}, \xi}-\frac{3-3\xi}{1+3\xi}\phi_\eta.
\end{equation}
When $\xi< 1/3$, we replace the input state with $\rho_{\text{w}, 1/3}$ and implement the above process. In each reputation of the experiment, we estimate the expectation value of observables to the state given by the quantum teleportation process. Besides, the accuracy of the estimated value can be systematically improved by repeating the experiments more times to reduce the statistical fluctuation. The expected number of repetitions is proportional to the cost specified in Eq.~\eqref{eq:ent_cost}. In each round of the experiments, only one copy of the input state is demanded, underscoring the strength of the VRD protocol.

\section{Experimental details}
We start by presenting an overview of the critical optical components utilized in our experimental setup and elucidate their respective functions:
\begin{enumerate}
    \item Waveplates: Central to our experimental design are HWPs and QWPs. The angle parameters, $\theta$ for HWPs and $\zeta$ for QWPs, define the orientation of the fast axis relative to the vertical polarization direction.
    
- HWP Unitary Transformation ($U_{\text{HWP}}(\theta)$):
    \[
    U_{\text{HWP}}(\theta) =- \begin{pmatrix}
    \cos 2\theta & \sin 2\theta \\
    \sin 2\theta & -\cos 2\theta \\
    \end{pmatrix}
    \]

    - QWP Unitary Transformation ($U_{\text{QWP}}(\zeta)$):
    \[
    U_{\text{QWP}}(\zeta) = \frac{1}{\sqrt{2}}\begin{pmatrix}
    1+i\cos 2\zeta & i\sin 2\zeta \\
    i\sin 2\zeta & 1-i\cos 2\zeta \\
    \end{pmatrix}
    \]
    \item Beam displacer~(BD): A BD transmits the vertically polarized photons and deviates the horizontally polarized photons.
    \item Polarization Beam Splitter~(PBS): A PBS transmits the horizontal polarization and reflects the vertical polarization.
    \item Beam Splitter~(BS): A BS transmits and reflects photons, either vertically polarized or horizontally polarized, with probability of 50\%.
\end{enumerate}

\subsection{Polarization-entangled photon source}
\begin{figure}[h!]
    \centering
    \includegraphics[width=0.4\linewidth]{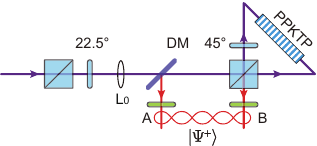}
    \caption{Schematic illustration of setup to generate polarization-entangled photon pair. }
    \label{fig:source}
\end{figure}
The experimental setup to generate polarization-entangled photon pairs are shown in Fig.~\ref{fig:source}. The entangled photons are generated from a type-II quasi-phase-matched Potassium Titanyl Phosphate~(PPKTP) crystal via spontaneous parametric down-conversion~(SPDC), which is set at the center of a Sagnac interferometer and bidirectionally pumped by an ultraviolet laser with central wavelength of 405~nm. A HWP set at $22.5^\circ$ rotates the polarization of pump light from $\ket{H}_P$ to $\ket{+}_\text{P}=\frac{1}{\sqrt{2}}(\ket{H}_\text{P}+\ket{V}_\text{P})$. Then, two lenses~(denoted as L$_0$ in Fig.~\ref{fig:source}) with focal length of 75~mm and 125~mm focus the pump beam at the center of PPKTP crystal with waist size of 74~$\mu$m. The pump light with horizontal polarization~($\ket{H}_\text{P}$) pumps PPKTP crystal from counterclockwise direction and generates photon pair in state of $\ket{HV}$. The vertical component~($\ket{V}_\text{P}$), which is rotated by a HWP set at 45$^\circ$, pumps PPKTP crystal from clockwise direction and generates photon pair in state of $\ket{HV}$. The photons from SPDC are with central wavelength of 810~nm, and superposed on the PBS, leading the output state in the ideal form of $\ket{\Psi^{+}}=\frac{1}{\sqrt{2}}(\ket{HV}+\ket{VH})$.

\subsection{Preparation and detection of single-ququart states}
\begin{figure}[h]
    \centering
\includegraphics[width=0.8\linewidth]{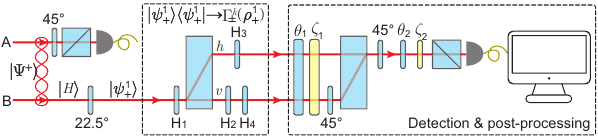}
    \caption{(a) The experimental setup to generate single-ququart states. (b) The experimental setup to measure single-ququart states.}
    \label{fig:ququart}
\end{figure}
As shown in Fig.~\ref{fig:ququart}, four HWPs H$_1\sim$H$_4$ with angles $\theta_1\sim\theta_4$  is able to transform $\ket{\psi_+^1}=(\ket{Hv}+\ket{Vv})/\sqrt{2}$ to $\rho_\pm^k=\Gamma(\rho_+^1)$.  For example,  the setting of angles $\bm{\theta}_+^4=[45^\circ, 0^\circ, 0^\circ, 0^\circ]$ can generate $\ket{\psi_+^4}$ by 
\begin{equation}\label{Eq:ququart}
\begin{aligned}
\ket{\psi_+^1}&=\frac{1}{\sqrt{2}}(\ket{Hv}+\ket{Vv})\\
&\xrightarrow{\theta_1=45^\circ} \frac{1}{\sqrt{2}}(\ket{Hv}+\ket{Vv})\\
&\xrightarrow{\text{BD}} \frac{1}{\sqrt{2}}(\ket{Hh}+\ket{Vv})\\
&\xrightarrow{\theta_2,\theta_3,\theta_4=0^\circ}\frac{1}{\sqrt{2}}(\ket{Hh}+\ket{Vv})=\frac{1}{\sqrt{2}}(\ket{1}+\ket{2})=\ket{\psi_+^4}.
\end{aligned}
\end{equation}
Recall the encoding role: $\ket{0}=\ket{H}\ket{v}$, $\ket{1}=\ket{V}\ket{v}$, $\ket{2}=\ket{H}\ket{h}$, and $\ket{3}=\ket{V}\ket{h}$. The setting of angles $\theta_1\sim\theta_4$ are listed in Tab.~\ref{tab:thetavalue}. Note that the setting of angles $\bm{\theta}_\pm^k=[\theta_{k_1}, \theta_{k_2}, \theta_{k_3}, \theta_{k_4}]$ is not unique. 
\begin{table*}[h!]
    \centering
    \begin{tabular}{c|cccccc}
    \hline\hline
       $k$ & $1$ & $2$ & $3$ & $4$ & $5$ & $6$ \\
       \hline
      $\bm{\theta}_k^+$&  [67.5$^\circ$, 0$^\circ$, 0$^\circ$, 22.5$^\circ$] & [45$^\circ$, 45$^\circ$, 0$^\circ$, 0$^\circ$] & [45$^\circ$, 45$^\circ$, 45$^\circ$, 0$^\circ$] & [45$^\circ$, 0$^\circ$, 0$^\circ$, 0$^\circ$] & [45$^\circ$, 0$^\circ$, 45$^\circ$, 0$^\circ$] & [22.5$^\circ$, 0$^\circ$, 22.5$^\circ$, 0$^\circ$]\\
      $\bm{\theta}_k^-$&  [67.5$^\circ$, 0$^\circ$, 0$^\circ$, 67.5$^\circ$] & [0$^\circ$, 45$^\circ$, 0$^\circ$, 0$^\circ$] & [0$^\circ$, 45$^\circ$, 45$^\circ$, 0$^\circ$] & [0$^\circ$, 0$^\circ$, 0$^\circ$, 0$^\circ$] & [0$^\circ$, 0$^\circ$, 45$^\circ$, 0$^\circ$] & [22.5$^\circ$, 0$^\circ$, 67.5$^\circ$, 0$^\circ$] \\
    \hline\hline
    \end{tabular}
    \caption{The setting of angles $\bm{\theta}_\pm^k$.}
    \label{tab:thetavalue}
\end{table*}

The projective measurements of prepared $\ket{\psi_\pm^k}$ on arbitrary basis $(\alpha \ket{H}+\delta \ket{V})\otimes(\gamma\ket{h}+\mu\ket{v})$ can be realized by the setup shown in Fig.~\ref{fig:ququart}. The projection on polarization DOF $\alpha \ket{H}+\delta \ket{V}$ is achieved by a HWP~(set at angle $\theta_1$) and a QWP~(set at angle $\zeta_1$), while the projection on path DOF $\gamma\ket{h}+\mu\ket{v}$ is realized with a HWP~(set at angle $\theta_2$), QWP~(set at angle $\zeta_2$) and a PBS. The step-by-step description of this projective measurement is 
\begin{equation}
\begin{aligned}
&(\alpha\ket{H}+\beta \ket{V})\otimes(\gamma\ket{h}+\delta\ket{v})\\
& \xrightarrow[\text{on both path}]{\text{HWP}@\theta _1\&\text{QWP}@\zeta_1} \ket{H}\otimes(\gamma\ket{h}+\delta\ket{v}) \\
& \xrightarrow[\text{on\,\,path\,\,}h]{\text{HWP}@45^\circ} \gamma\ket{Vh}+\delta\ket{Hv} \\
& \xrightarrow[\text{combine the path}\,\,h\,\,\text{and}\,\,v]{\rm{BD3}} \gamma\ket{V}+\delta\ket{H} \\
& \xrightarrow{\rm{HWP@45^\circ}} \gamma\ket{H}+\delta\ket{V} \\
& \xrightarrow[]{\text{HWP}@\theta_2\&\text{QWP}@\zeta_2} \ket{H} \\ 
\end{aligned}.
\end{equation}

In our experiment, we perform the Pauli measurements, and the corresponding setting of angels $\theta_1$, $\zeta_1$, $\theta_2$ and $\zeta_2$ are listed in Tab.~\ref{tab:Paulisetting}.
\begin{table*}[htb]
    \centering
    \begin{tabular}{c|c|c|c|c|c|c|c|c|c}
    \hline\hline
         & $\ket{H}$ & $\ket{V}$ & $(\ket{H}+\ket{V})/\sqrt{2}$ & $(\ket{H}+i\ket{V})/\sqrt{2}$ && $\ket{h}$ & $\ket{v}$ & $(\ket{h}+\ket{v})/\sqrt{2}$ & $(\ket{h}+i\ket{v})/\sqrt{2}$  \\\hline
        $\theta_1$ &0$^\circ$&45$^\circ$&22.5$^\circ$&0$^\circ$&$\theta_2$&0$^\circ$&45$^\circ$&22.5$^\circ$&0$^\circ$\\
        $\zeta_1$ &0$^\circ$&0$^\circ$&0$^\circ$&45$^\circ$&$\zeta2$&0$^\circ$&0$^\circ$&0$^\circ$&45$^\circ$\\
       \hline\hline
    \end{tabular}
    \caption{The setting of angles $\bm{\theta}_\pm^k$.}
    \label{tab:Paulisetting}
\end{table*}

\subsection{Generation of $\rho_{\text{w}, \xi}$ and $\rho_\eta$}
The Werner state can be decomposed into a mixture of four pure states
\begin{equation}\label{Eq:WSexp}
\rho_{\text{w}, \xi}=\frac{1+3\xi}{4}\Psi^-+\frac{1-\xi}{4}\left(\Psi^++\rho_{HH}+\rho_{VV}\right). 
\end{equation}
\begin{figure}[htb]
    \centering
    \includegraphics[width=0.6\linewidth]{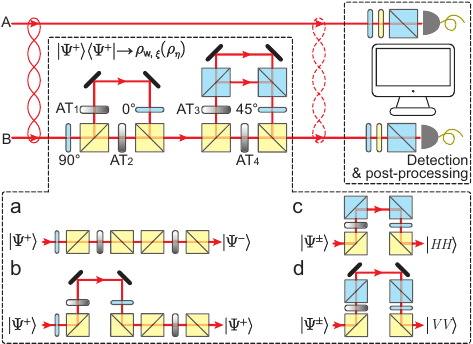}
    \caption{Experimental setup to generate $\rho_{\text{w},\xi}$ and $\rho_\eta$.}
    \label{fig:werner}
\end{figure}
As shown in Fig.~\ref{fig:werner}(a)-(d), $\ket{\Psi^+}$ can be converted to $\ket{\Psi^+}$, $\ket{\Psi^-}$, $\ket{HH}$ and $\ket{VV}$, respectively. Specifically, $\ket{\Psi^+}$ is converted to $\ket{\Psi^-}$ by a HWP set at 90$^\circ$ with probability of $t_{\xi_2}t_{\xi_4}/16$ as shown in Fig.~\ref{fig:werner}(a), where the factor $1/16$ comes from four BS and $t_{\xi_2}$ and $t_{\xi_4}$ are the transmittance of AT$_2$ and AT$_4$. As shown in Fig.~\ref{fig:werner}(b), $\ket{\Psi^+}$ is attenuated by probably of $t_{\xi_1}t_{\xi_4}/16$, where $t_{\xi_1}$ is the transmittance of AT$_1$. As shown in  Fig.~\ref{fig:werner}(c)-(d), either $\ket{\Psi^+}$ or $\ket{\Psi^-}$ is converted to $\ket{HH}$~($\ket{VV}$) with probability of $t_{\xi_3}/8$,  where $t_{\xi_3}$ is the transmittance of AT$_3$. Multiplied by the attenuation $(t_{\xi_1}+t_{\xi_2})/4$ of $\ket{\Psi^\pm}$, the initial $\ket{\Psi^+}$ is converted to $\ket{HH}$~($\ket{VV}$) with probability of $(t_{\xi_1}+t_{\xi_2})t_{\xi_3}/32$. The four states are incoherently mixed on a BS, leading the normalized outcome state of   
\begin{equation}\label{Eq:Wderive}
\frac{1}{(t_{\xi_1}+t_{\xi_2})(t_{\xi_3}+t_{\xi_4})}\left[t_{\xi_2}t_{\xi_4}\Psi^-+t_{\xi_1}t_{\xi_4}\Psi^++\frac{(t_{\xi_1}+t_{\xi_2})t_{\xi_3}}{2}(\rho_{HH}+\rho_{VV})\right]
\end{equation}
Compare Eq.~\eqref{Eq:Wderive} and Eq.~\eqref{Eq:WSexp}, it is easy to obtain the setting of transmittance $\bm{t}(\xi)=[t_1, t_2, t_3,t_4]$ to prepare $\rho_{\text{w}, \xi}$
\begin{equation}
t_{\xi_1}=\frac{1-\xi}{2+2\xi},
t_{\xi_2}=\frac{1+3\xi}{2+2\xi},
t_{\xi_3}=\frac{1-\xi}{2},
t_{\xi_4}=\frac{1+\xi}{2}. 
\end{equation}
Similarly, $\rho_\eta = \frac{1}{3}\left(\Psi^+ + \rho_{HH} + \rho_{VV}\right)$ can be generated with the same setup by setting $t_{\eta_1}=1$, $t_{\eta_2}=0$, $t_{\eta_3}=\frac{2}{3}$ and $t_{\eta_4}=\frac{1}{3}$.

\subsection{More results of quantum teleportation}
\begin{figure*}[h]
    \centering
    \includegraphics{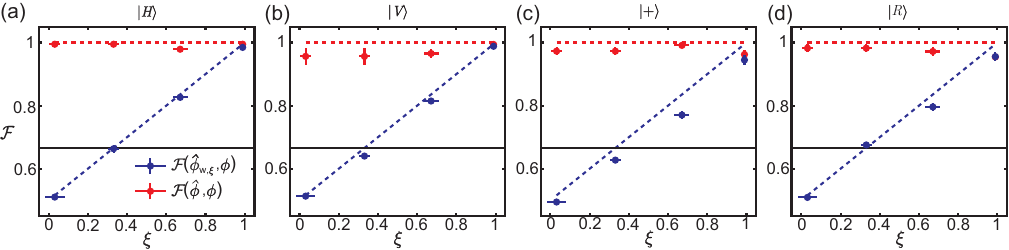}
    \caption{The fidelity of the teleported state with respect to their input state $\ket{\phi}\in\left\{ \ket{H}, \ket{V}, \ket{+}, \ket{R} \right\} $ based on Werner states before and after distillation. The black dashed line represent classical threshold $fc = 2/3$ in quantum teleportation.}
    \label{fig:allstate}
\end{figure*}

Step-by-step descriptions of two-photon teleportation are presented in Appendix C of Ref.~\cite{LiJyun-Yi21}. We show the fidelity $\mathcal F(\hat{\phi}, \phi)$ with $\ket{\phi}\in[\ket{H}, \ket{V}, \ket{+}, \ket{R}]$ in Fig.~\ref{fig:allstate}(a)-(d). 

\subsection{Estimation of unknown quantum states }
The complete estimation of unknown quantum states requires information-complete measurements such as Pauli measurements, i.e., the projective measurements on eigenvectors of Pauli operator $X$, $Y$ and $Z$. In the standard Pauli basis $\{\mathds 1, X, Y, Z\}$, a $n$-qubit state $\rho$ can be expanded as
\begin{equation}
\rho=\frac{1}{2^n}\sum_{v}\tr(M_{v}\rho)M_{v},
\end{equation}  
where $v=1, 2, \cdots 4^n$ enumerates all possible $n$-fold tensor products of Pauli matrices $M_1=\mathds 1\otimes\mathds 1\otimes\cdots\otimes\mathds1$,  $M_2=\mathds 1\otimes\mathds 1\otimes\cdots\otimes X$, $\dots$, $M_{4^n}=Z\otimes Z\otimes\cdots\otimes Z$. Experimentally, we are able to detect the expected values of observable $M_v$, denoted as $f_v$. Then the estimation of $\rho$ with linear inversion~(LIN) is 
\begin{equation}
\hat{\rho}^\text{LIN}=\frac{1}{2^n}\sum_{v}f_vM_{v}.
\end{equation}
With finite number of experimental runs, $f_v\neq\tr(M_{v}\rho)$ so that $\hat{\rho}^\text{LIN}$ is generally not a physical~(quantum) state. This can be addressed by solving the optimization 
\begin{equation}
\label{Eq:opt}
\hat{\rho}=\mathop{\text{argmin}}\limits_ {\sigma\geq0, \tr(\sigma)=1}\hspace{0.5cm}||\sigma-\hat{\rho}^\text{LIN}||_2,
\end{equation}
where $||\cdot||_2$ is the Frobenius distance.

\section{Limitations of virtual distillation in retrieving information}

Virtual distillation is only a very `simple', that is linear, way to post-process measurement statistics. From the most fundamental perspective of information processing, there should exist more efficient ways to distil states and retrieve information.

Here we give an explicit example showing that virtual distillation has no advantage in retrieving information.
Consider an input quantum state $\rho$, suppose a transformation $\exp (-i \mathbb{A} \theta)$ with a parameter $\theta$ is applied to $\rho$, we have an output state
$$
\rho(\theta)=\exp (-i \mathbb{A} \theta) \rho \exp (+i \mathbb{A} \theta).
$$
The quantum Fisher information $F_{\mathrm{Q}}[\rho, \mathbb{A}]$ of $\rho$ with respect to the observable $\mathbb{A}$ is defined as
$$
F_{\mathrm{Q}}[\rho, \mathbb{A}]=2 \sum_{k, l} \frac{\left(\lambda_k-\lambda_l\right)^2}{\left(\lambda_k+\lambda_l\right)}|\langle k|\mathbb{A}| l\rangle|^2,
$$
where $\lambda_k$ and $|k\rangle$ are the eigenvalues and eigenvectors of the density matrix $\rho$, respectively, and the summation goes over all $k$ and $l$ such that $\lambda_k+\lambda_l>0$.
The quantum Fisher information constrains the achievable precision in statistical estimation of the parameter $\theta$ via the quantum Cram\'{e}r-Rao bound as
$$
\Delta \theta \geq \frac{1}{\sqrt{m F_{\mathrm{Q}}[\rho, \mathbb{A}]}},
$$
where $m$ is the number of independent repetitions.

As an example, here we consider a noisy entangled state $\rho_{AB}(\xi)=\xi \ket{\Psi_{AB}}\bra{\Psi_{AB}} + (1-\xi)\mathds1_4/4$ as the input state with observable $\mathbb{A}=Z_A + Z_B$. 
Here $\ket{\Psi}_{AB}=(\ket{00}_{AB}+\ket{11}_{AB})/\sqrt{2}$. We can calculate the quantum Fisher information of $\rho_{AB}$ and $\ket{\Psi_{AB}}\bra{\Psi_{AB}}$ as
\begin{equation}
    \begin{aligned}
    F_{\mathrm{Q}}[\rho_{AB}(\xi), \mathbb{A}] &= \frac{8\xi^2}{(1+\xi)}\bra{\Psi_{AB}} \mathbb{A} [I_4 - \ket{\Psi_{AB}}\bra{\Psi_{AB}}] \mathbb{A} \ket{\Psi_{AB}}=\frac{32\xi^2}{(1+\xi)}\\
        F_{\mathrm{Q}}[\ket{\Psi_{AB}}\bra{\Psi_{AB}}, \mathbb{A}] &= 4\bra{\Psi_{AB}} \mathbb{A} [I_4 - \ket{\Psi_{AB}}\bra{\Psi_{AB}}] \mathbb{A} \ket{\Psi_{AB}}
        = 16
    \end{aligned}
\end{equation}
where we have used $\bra{\Psi_{AB}} \mathbb{A} [I_4 - \ket{\Psi_{AB}}\bra{\Psi_{AB}}] \mathbb{A} \ket{\Psi_{AB}}=\bra{\Psi_{AB}} \mathbb{A}^2 \ket{\Psi_{AB}} - \bra{\Psi_{AB}} \mathbb{A} \ket{\Psi_{AB}}^2=4$.

\begin{figure}
    \centering\includegraphics[width=0.5\textwidth]{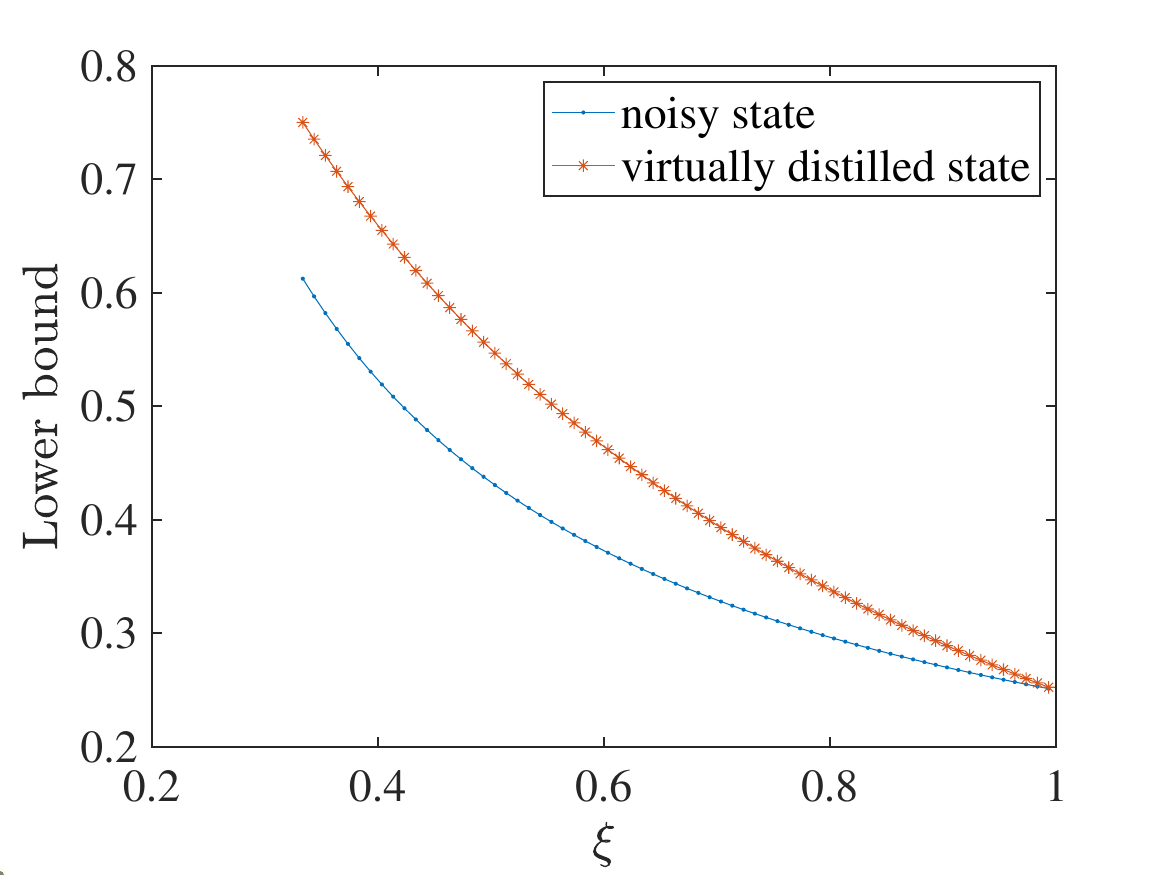}
\caption{Comparison of the quantum Cram\'{e}r-Rao bound for the noisy and virtually distilled states with different noise ratio $\xi$.}
    \label{fig:enter-label}
\end{figure}

Now consider $m$ copies of the noisy entangled state $\rho_{AB}(\xi)$, the quantum Cram\'{e}r-Rao bound is 
$$
\Delta \theta \geq \frac{1}{\sqrt{m F_{\mathrm{Q}}[\rho_{AB}(\xi), \mathbb{A}]}} = \sqrt{\frac{1+\xi}{32\xi^2}}\frac{1}{\sqrt{m}},
$$
On the other hand, suppose we apply virtual resource distillation on $\rho_{AB}(\xi)$.
Suppose $\xi\ge 1/3$, the virtual distillation rate is $V^0(\rho_{AB}(\xi)) = 1/C^0(\rho_{AB}(\xi))^2 = (1+3\xi)^2/(7-3\xi)^2$.
Then the 
quantum Cram\'{e}r-Rao bound is 
$$
\Delta \theta \geq \frac{1}{\sqrt{mV^0(\rho_{AB}(\xi)) F_{\mathrm{Q}}[\psi_{AB}, \mathbb{A}]}} = \sqrt{\frac{(7-3\xi)^2}{16 (1+3\xi)^2}}\frac{1}{\sqrt{m}},
$$
We compare the coefficients $\sqrt{\frac{1+\xi}{32\xi^2}}$ versus $\sqrt{\frac{(7-3\xi)^2}{16 (1+3\xi)^2}}$ for the noisy entangled state and the virtually distilled state as shown in Fig.~\ref{fig:enter-label}.
We can see that the lower bound of the virtually distilled state is strictly larger than the one of the noisy state for $\xi \neq 1$, which indicates that the virtually distilled state is worse at retrieving information. 

At last, we would like to clarify again that virtual distillation is not designed as a better way to retrieve information. Instead, virtual distillation is proposed as a better and analytically tractable way of distilling quantum resources, which plays an important role in quantum communication and computation. If we only aim at retrieving the information of a noisy state, we would like to conjecture that any quantum operation + post-processing cannot increase the information gain. Nevertheless, the lower bound might also be hard to reach, which might generally require complicated post-processing of the measurement statistics.

\end{document}